\documentclass[aps,prl,showpacs,amsmath,amssymb,amsfonts,twocolumn,
superscriptaddress,floatfix,footinbib]{revtex4}
\usepackage{psfig}
\usepackage{graphicx}

\newcommand{\bgar}{\begin{eqnarray}}
\newcommand{\enar}{\end{eqnarray}}
\newcommand{\eq}[1]{(\ref{eq:#1})}

\newcommand{\eqname}[1]{\label{eq:#1}}

\newcommand{\kk}{ {\bf k}}
\newcommand{\vv}{ {\bf v}}
\newcommand{\xx}{ {\bf x}}

\newcommand{\Psihd}{\hat\Psi^\dagger}

\newcommand{\Psih}{\hat\Psi}

\newcommand{\Hamilt}{{\mathcal H}}

\begin{document}

\title{Probing microcavity polariton superfluidity through resonant Rayleigh
scattering}

 \author{I. Carusotto}
\email{Iacopo.Carusotto@lkb.ens.fr}
 \affiliation{Laboratoire Kastler Brossel, \'Ecole Normale Sup\'erieure, 24 rue
 Lhomond, F-75231 Paris Cedex 05, France}
\affiliation{CRS BEC-INFM and Dipartimento di Fisica, Universit\`a di
  Trento, I-38050 Povo, Italy}

\author{C. Ciuti}
 \affiliation{Laboratoire Pierre Aigrain, \'Ecole Normale Sup\'erieure, 24 rue
 Lhomond, F-75231 Paris Cedex 05, France}

\pacs{71.36.+c, 42.65.-k, 03.75.Kk}
 \date{\today}
%71.36.+c Polaritons (including photon-phonon and photon-magnon
% interactions)
%42.50.-p Quantum optics
%42.65.-k Nonlinear optics
%71.35.Lk Collective effects (Bose effects, phase space filling, and
% excitonic phase transitions)
% 03.75.Kk Dynamic properties of condensates; collective and
% hydrodynamic excitations, superfluid flow

\begin{abstract}

We investigate the two-dimensional motion of polaritons injected into a 
planar microcavity by a continuous wave optical pump in presence of
a static perturbation, e.g. a point defect. By finding the stationary
solutions of the nonlinear mean-field equations (away
from any parametric instability), we show how the spectrum of the
polariton Bogoliubov-like excitations reflects onto the shape and
intensity of the resonant Rayleigh scattering emission pattern in both
momentum and real space. 
We find a superfluid regime in the sense of the Landau criterion, in which 
the Rayleigh scattering ring in momentum space collapses as well as
its normalized intensity. More generally, we show how collective
excitation spectra having no analog in equilibrium systems can be
observed by tuning the excitation angle and frequency. Predictions
with realistic semiconductor microcavity parameters are given.

\end{abstract}

 \maketitle

%\section{Introduction}

The concept of a quantum fluid has played a central role in many
fields of condensed matter and atomic physics, ranging from
superconductors to Helium fluids~\cite{ManyBody} and, more recently, atomic
Bose-Einstein condensates~\cite{AtomicBEC}. One of the most
exciting manifestations of quantum behavior is superfluidity,
i.e. the possibility of the fluid to flow without friction around
an impurity~\cite{Superfluidity}.

In this Letter, we investigate the superfluid properties of a
two-dimensional gas of polaritons in a semiconductor microcavity
in the strong light-matter coupling regime \cite{Weisbuch}. In
this system, the normal modes are superpositions of a cavity
photon and a quantum well exciton. Thanks to their photonic
component, polaritons can be coherently excited by an incident
laser field and detected through angularly or spatially resolved
optical spectroscopy. Thanks to their excitonic component,
polaritons have strong binary interactions, which have been
demonstrated to produce spectacular polariton amplification
effects \cite{Savvidis,Saba} through matter-wave stimulated
collisions \cite{Ciuti}, as well as spontaneous parametric
instabilities~\cite{Baumberg,Messin,Review}.

Here, we study the propagation of a polariton fluid in presence of
static imperfections, which are known to produce resonant Rayleigh
scattering (RRS) of the exciting laser
field~\cite{RRS,HoudreRRSLin,HoudreRRSNLin,Langbein_ring}. We show
that superfluidity of the polariton fluid manifests itself as a
quenching of the RRS intensity when the flow velocity imprinted by
the exciting laser is slower than the sound velocity in the
polariton fluid. Furthermore, a dramatic reshaping of the RRS
pattern  due to polariton-polariton interactions can be observed
in both momentum and real space even at higher flow velocities.
Interestingly, the polariton field oscillation frequency is not
fixed by an equation of state relating the chemical potential to
the particle density, but it can be tuned by the frequency of the
exciting laser. This opens the possibility of having a spectrum of
collective excitation around the stationary state which has no
analog in usual systems close to thermal equilibrium. We show in
detail how these peculiar excitation spectra can be probed by
resonant Rayleigh scattering.

\begin{figure}[htbp]
\begin{center}
\includegraphics[width=8cm,clip]{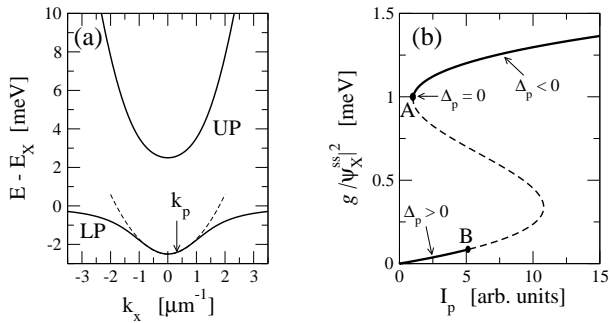}
\caption{(a): linear dispersion of the lower (LP) and upper (UP)
polariton branches. Typical position of the pump wavevector
$\kk_p$ (arrow), well in the parabolic region of the LP
dispersion; the dashed line is the parabolic approximation at
small $\kk$. (b): internal vs. incident intensity curve showing
bistable behavior. The dashed branch is unstable: the inversion
points $A$ and $B$ are  respectively due to a single-mode (Kerr)
or a multi-mode (parametric) instability. Excitation parameters:
$k_p=0.314\,\mu\textrm{m}^{-1}$,
$\omega_p-\omega_{LP}(\kk_p)=0.47\,\textrm{meV}$. Cavity
parameters are taken from ref.\cite{HoudreRRSLin}:
$\gamma=0.1\,\textrm{meV}$,
$\omega_X=\omega_C^0=1.4\,\textrm{eV}$,
$2\,\Omega_R=5\,\textrm{meV}$. \label{fig:figura1}}
\end{center}
\end{figure}

A commonly used model for describing a planar microcavity
containing a quantum well with an excitonic resonance strongly
coupled to a cavity mode is based on the Hamiltonian
\cite{Ciuti_Review}:
\begin{multline}
  \label{eq:Hamilt_tot}
  \Hamilt=\int\!d\xx\,\sum_{ij=\{X,C\}}
\Psihd_{i}(\xx) \,\Big[{\mathbf
  h}^{0}_{ij}+V_{i}(\xx)\,\delta_{ij} \Big]\,\Psih_{j}(\xx)\\
%+\sum_{i} V_{i}(\xx)\,\Psihd_{i}(\xx)\,\Psih_{i}(\xx) \\
+\frac{\hbar g}{2}\int\! d\xx\,\Psihd_{X}(\xx)\,\Psihd_{X}(\xx)\,
\Psih_{X}(\xx)\,\Psih_{X}(\xx)+\\
+\int\!d\xx\,\hbar F_{p}\,e^{i(\kk_{p} \xx-\omega_{p}t)}
\,\Psihd_{C}(\xx)+\textrm{h.c.}~,
\end{multline}
where $\xx$ is the in-plane spatial position and the field
operators $\Psi_{X,C}(\xx)$ respectively describe excitons ($X$)
and cavity photons ($C$). They satisfy Bose commutation rules,
$[\Psih_i(\xx),\Psihd_j(\xx')]=\delta^2(\xx-\xx')\,\delta_{ij}$.
The linear Hamiltonian ${\mathbf h}^0$ is:
\begin{equation}
  \label{eq:Hamilt_lin}
  {\mathbf h}^0=
\hbar \left(
\begin{array}{cc}
\omega_{X}(-i\nabla) & \Omega_R \\
\Omega_R & \omega_C(-i\nabla)
\end{array}
\right),
\end{equation}
where $\omega_{C}(\kk)=\omega_{C}^0\,\sqrt{1+{\kk^2}/{k_z^2}}$ is
the cavity dispersion as a function of the in-plane wavevector
$\kk$ and $k_z$ is the quantized photon wavevector in the growth
direction. $\Omega_R$ is the Rabi frequency of the exciton-cavity
photon coupling. A flat exciton dispersion
$\omega_X(\kk)=\omega_X$ will be assumed in the following. In this
framework, polaritons simply arise as the eigenmodes of the linear
Hamiltonian \eq{Hamilt_lin}; $\omega_{LP(UP)}(\kk)$ denotes the
dispersion of the lower (upper) polariton branch
[Fig.\ref{fig:figura1}(a)].

The external force term proportional to $F_{p}$ describes a
coherent
and monochromatic laser field of frequency $\omega_{p}$ (called
the {\em pump}), which drives the cavity and injects polaritons.
Spatially, it is assumed to have a plane-wave profile of
wavevector $k_{p}=\sin\theta_p\; \omega_{p}/c$, $\theta_p$ being
the pump incidence angle, so to generate a polariton fluid with a
non-zero flow velocity along the cavity plane.
The nonlinear interaction term is due exciton-exciton collisional
interactions and, as usual, is modelled by a repulsive ($g>0$) contact
potential. The anharmonic exciton-photon coupling has a negligible
effect in the regime considered in the present study~\cite{Ciuti_Review}.
$V_{X,C}(\xx)$ are external potential terms acting on the excitonic
and photonic fields which can model the presence of disorder.
Here, results for the specific case of a point defect will be
presented. Note that point defects can be naturally present in
state-of-the-art
samples~\cite{Langbein} or even be created deliberately by means
of lithographic techniques.

Within the mean-field approximation, the time-evolution of the mean fields
$\psi_{X,C}(\xx)=\langle\Psih_{X,C}(\xx) \rangle$ under the Hamiltonian
\eq{Hamilt_tot} is given by:
\begin{multline}
  \label{eq:GPE}
  i\,\frac{d}{dt}
\left(
  \begin{array}{c}
\psi_{X}(\xx) \\ \psi_{C}(\xx)
  \end{array}
\right)=
\left(
  \begin{array}{c}
0 \\ F_{p} \, e^{i(\kk_p\xx-\omega_p t)}
  \end{array}
\right)+
\left[
{\mathbf h}^0-\frac{i\gamma}{2}\,{\mathbf 1}+\right. \\
+\left.\left(
\begin{array}{cc}
V_X(\xx)+g|\psi_X(\xx)|^2 & 0 \\ 0 & V_C(\xx)
  \end{array}
\right)
\right]
\left(
  \begin{array}{c}
\psi_{X}(\xx) \\ \psi_{C}(\xx)
  \end{array}
\right)
.
\end{multline}
In the quantum fluid language, these are the Gross-Pitaevskii
equations~\cite{AtomicBEC} for our cavity-polariton system. For
simplicity, an equal rate $\gamma$ is assumed for the damping of
both the excitonic and the photonic fields. In the present work,
we will be concerned with an excitation close to the bottom of the
LP dispersion, i.e. the region most protected~\cite{Review} from
the exciton reservoir, which may be responsible for
excitation-induced decoherence~\cite{savasta}.

%\section{The stationary state: bistability effects}

In the homogeneous case ($V_{X,C}=0$), we can look for
spatially homogeneous stationary states of the system in which the
field has the same plane wave structure
$\psi_{X,C}(\xx,t)=\exp[i(\kk_{p} \xx-\omega_{p}
t)]\,\psi^{ss}_{X,C}$ as the incident laser field. The resulting
equations
\begin{eqnarray}
\Big(\omega_{X}(\kk_{p})-\omega_{p}-\frac{i}{2}\gamma+
g\,|\psi_X^{ss}|^2\Big)\,
\psi^{ss}_{X}+\Omega_R\,\psi_C^{ss}=0  \label{eq:ss_X} \\
\Big(\omega_{C}(\kk_{p})-\omega_{p}-\frac{i}{2}\gamma \Big)\,
\psi^{ss}_{C}+\Omega_R\,\psi_X^{ss}=-F_p,  \label{eq:ss_C}
\end{eqnarray}
are the generalization of the state equation. While the
oscillation frequency of the condensate wavefunction  in an
isolated gas is equal to the chemical potential $\mu$ and
therefore it is fixed by the equation of state, in the present
driven-dissipative system it is equal to the frequency
$\omega_{p}$ of the driving laser and therefore it is an
experimentally tunable parameter. As usual, stability of the
solutions of Eqs. (\ref{eq:ss_X}-\ref{eq:ss_C}) has to be checked
by linearizing Eq. \eq{GPE} around the stationary state.
For $\omega_p>\omega_{LP}(\kk_p)$, the relation between
the incident intensity $I_p\propto |F_{p}|^2$ and the internal one
shows the typical S-shape of optical bistability (see
Fig.\ref{fig:figura1}b and
Refs.\cite{BistableExp,BistableExp2,Bistability}). Note that nice
hysteresis loops due to polariton bistability have been recently
experimentally demonstrated \cite{BistableExp} in the case $\kk_p
= 0$.
In the opposite case $\omega_p<\omega_{LP}(\kk_p)$ (not shown),
the behaviour of the
system would instead be the typical one of an optical
limiter~\cite{Bistability}.

In the stability region, the response of the system to a weak
perturbation can be determined by means of the linearized theory.
In the field of quantum fluids, this approach is called Bogoliubov
theory~\cite{AtomicBEC}. By defining the slowly varying fields
with respect to the pump frequency as $\delta\phi_{i}(\xx,t) =
\delta {\psi_{i}}(\xx,t) \exp(i\omega_p t)$, the motion equation
of the four-component  displacement vector $\delta {\vec
\phi}(\xx,t)= \big( \delta\phi_{X}(\xx,t), \delta\phi_{C}(\xx,t),
\delta\phi^*_{X}(\xx,t), \delta\phi_{C}^*(\xx,t) \big)^T$ reads
\begin{equation}
  \label{eq:Bogo_motion}
i\frac{d}{dt} \delta{\vec \phi}={\mathcal L}\cdot \delta{\vec
\phi}+{\vec f}_d~,
 \end{equation}
${\vec f}_d$ being the source term due to the weak perturbation
and the spectral operator ${\mathcal L}$ being defined as
\begin{widetext}
\begin{equation}
  \label{eq:Bogo_L}
  {\mathcal L}=
\left(
\begin{array}{cccc}
\omega_X+2g\,|\psi_X^{ss}|^2-  \omega_p -\frac{i\gamma}{2} &
\Omega_R & g\,\psi^{ss\,2}_X\,e^{2 i \kk_p\xx} & 0 \\
\Omega_R & \omega_C(-i\nabla)-  \omega_p - \frac{i\gamma}{2} & 0 & 0 \\
-g\,\psi^{ss\,*\,2}_{X}\,e^{-2 i \kk_p\xx} & 0 &
-\big(\omega_X+2g\,|\psi_X^{ss}|^2\big) +
\omega_p-\frac{i\gamma}{2} &
-\Omega_R \\
0 & 0 & -\Omega_R & -\omega_C(-i\nabla)+ \omega_p-\frac{i\gamma}{2}
\end{array}
\right).
\end{equation}
\end{widetext}

%\begin{widetext}

\begin{figure}[htbp]
\begin{center}
\includegraphics[width=8.5cm]{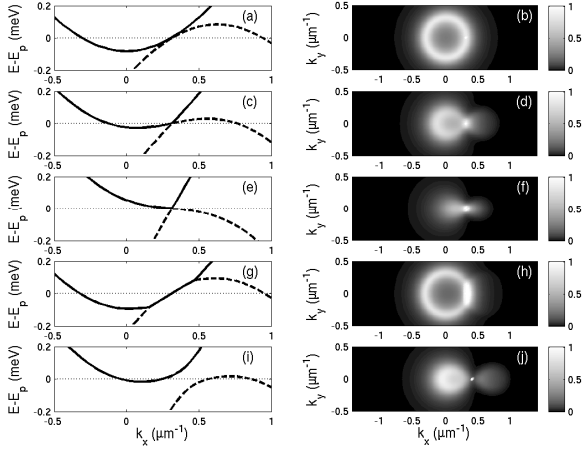}
\caption{Left panels: exact Bogoliubov dispersion for the LP
  branches calculated from Eq.\eq{Bogo_L}.  Right panels: corresponding
  RRS emission pattern in $\kk$-space. The intensity has been normalized to
  the transmitted intensity. The $\kk_p$ point
  (indicated by a white circle) saturates by far the gray scale.
Panels (a-f): resonant
  case $\Delta_p=0$
respectively in the linear regime (a-b), with
$g|\Psi^{ss}_X|^2=0.2\,\textrm{meV}$ (c-d), $1\,\textrm{meV}$
(e-f). Panels (g-h): $\Delta_p>0$ case with
  $g|\Psi^{ss}_X|^2=0.04\,\textrm{meV}$.
Panels (i-j): $\Delta_p <0$ case with $g|\Psi^{ss}_X|^2
=0.6\,\textrm{meV}$. Pump wavevector:
$k_p=0.314\,\mu\textrm{m}^{-1}$ (a-h), $0.408\,\mu\textrm{m}^{-1}$
(i-j). Same cavity parameters as in Fig.\ref{fig:figura1}.
\label{fig:comb_coll_k}}
\end{center}
\end{figure}
%\end{widetext}

The eigenvalues of the operator ${\mathcal L}$ correspond to the
frequencies of the Bogoliubov eigenmodes of the system. For each
$\kk$, the spectrum is composed of four branches
$\omega^{\pm}_{UP,LP}(\kk)$: for each polariton branch (LP or UP),
two $\pm$ branches exist, which are the image of each other under
the simultaneous transformations $\kk\rightarrow 2\kk_{p}-\kk$ and
$\omega\rightarrow 2\omega_{p}-\omega$~\cite{Ciuti_PL}. Numerical
calculations are shown in Fig.\ref{fig:comb_coll_k}. For the sake
of clarity, only the branches relative to the LP have been traced,
the ones relative to the UP being far away on the scale of the
figure. These numerical results can be understood through the
simplified analytical approximation that follows.

Provided the interaction energy $g\,|\psi_{X}^{ss}|^2$ is much
smaller than the polaritonic splitting $\omega_{UP}-\omega_{LP}$,
there is no significant mixing between the LP and UP branches.
Since we are interested in nearly-resonant excitation close to the
bottom of the LP dispersion curve,  we can describe the system in
terms of the LP field $\psi_{LP} = X_{LP} \psi_{X} + C_{LP}
\psi_{C}$ only, being $X_{LP}$ and $C_{LP}$ the Hopfield
coefficients quantifying the excitonic and photonic components. In
the parabolic approximation,
$\omega_{LP}(\kk)\simeq\omega_{LP}^0+\hbar\,\kk^2/ 2m_{LP}$ and
the self-coupling constant is $g_{LP}=g\,|X_{LP}|^4$. The
mean-field shift of the polariton mode is then
$\delta\omega_{MF}=g_{LP}\,|\psi_{LP}^{ss}|^2$. Under these
assumptions, the spectrum of the LP Bogoliubov excitations can be
approximated by the simple expression
\begin{equation}
\omega^\pm_{LP}\simeq \omega_p+ \delta\kk\cdot
\vv_p-\frac{i\gamma}{2}
\pm\sqrt{(2\,\delta\omega_{MF}+\eta_{\delta\kk}-\Delta_p)
(\eta_{\delta\kk}-\Delta_p)},
\eqname{Bogo}
\end{equation}
where $\delta\kk=\kk-\kk_{p}$,
$\eta_{\delta\kk}=\hbar\,\delta\kk^2/2m_{LP}$, the flow velocity
$\vv_p=\hbar\kk_p/m_{LP}$, and the effective pump detuning
\begin{equation}
\Delta_p=\omega_{p}-\omega_{LP}(\kk_p)-\delta\omega_{MF}.
\end{equation}

In the resonant case ($\Delta_p=0$), the $\pm$ branches touch at
$\kk=\kk_{p}$. The effect of the finite flow velocity $\vv_p$ is to
tilt the standard Bogoliubov dispersion~\cite{AtomicBEC} via the term
  $\delta\kk\cdot\vv_p$.
  While in the non-interacting case in Fig.\ref{fig:comb_coll_k}(a) the
  dispersion remains parabolic, in
  the presence of interactions [Fig.\ref{fig:comb_coll_k}(c) and (e)] its
  slope has a
  discontinuity at $\kk=\kk_{p}$: on each side of the corner,
  the $+$ branch starts linearly with group velocities
  respectively given by $v_g^{r,l}=c_s\pm v_p$, $c_s$ being the usual
  sound velocity of the
  interacting Bose gas $c_s=\sqrt{\hbar\,\delta\omega_{MF}/m_{LP}}$.
On the hysteresis curve of Fig.\ref{fig:figura1}(b), the condition
  $\Delta_p=0$ corresponds to the inversion point $A$.
If one moves to the right of the point $A$ along the upper branch of the
hysteresis curve, the mean-field shift
$\delta\omega_{MF}$ increases and the effective pump detuning
$\Delta_p$ becomes negative. 
In this case, as it is shown in Fig.\ref{fig:figura1}(i), the branches no
longer touch 
each other at $\kk_p$ and a full gap between them opens up for
sufficiently large values of $|\Delta_p|$ (not shown).

On the other hand, the effective pump detuning $\Delta_p$ is
strictly positive on the lower branch of the bistability curve of
Fig.\ref{fig:figura1}(b). In this case, the argument of the square
root in
  \eq{Bogo} is negative for the wavevectors $\kk$ such that
  $\Delta_p>\eta_{\delta\kk}>\Delta_p-2\,\delta\omega_{MF}$. In this
  region, the $\pm$ branches stick together~\cite{Ciuti_Review}
  (i.e. $\textrm{Re}[\omega_+] = \textrm{Re}[\omega_-]$) and have an
  exactly linear dispersion of slope $\vv_p$
  (Fig.\ref{fig:comb_coll_k}g). The imaginary parts are instead split, with
  one branch being narrowed and the other broadened~\cite{Ciuti_Review,Ciuti_PL}.
For
  $\delta\omega_{MF}>\gamma/2$, that is on the right of point $B$ in
  Fig.\ref{fig:figura1}(b), the multi-mode parametric
  instability~\cite{Ciuti_PL}  sets in.
In the field of quantum fluids, this kind of dynamical
instabilities are generally known as {\em modulational
instabilities}~\cite{NiuModInst}.

%\section{Effect of a static defect}

\begin{figure}[htbp]
\begin{center}
\includegraphics[width=8.5cm]{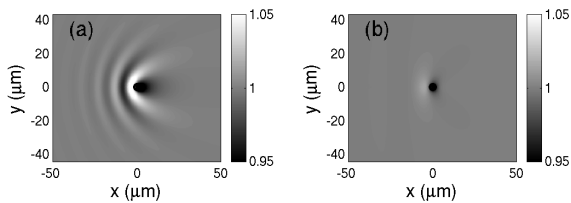}
\caption{ Real space resonant Rayleigh scattering emission pattern
for a localized defect at $x=y=0$ acting on the cavity photon. The
defect potential has a lateral size of $0.8\,\mu\textrm{m}$ and a
depth of $1\,\textrm{meV}$. In panel (a), linear regime as in
Fig.\ref{fig:comb_coll_k}(a) and (b). In panel (b), superfluid
regime as in Fig.\ref{fig:comb_coll_k}(e) and (f). In each panel,
the emitted intensity normalized to the transmitted intensity.
\label{fig:spazio_x}}
\end{center}
\end{figure}

The dispersion of the elementary excitations of the system is the
starting point for a study of its response to an external
perturbation. In particular, we shall consider here a weak and
static disorder as described by the potential $V_{C,X}(\xx)$. In
this case the perturbation source term ${\vec f}_{d}$ is
time-independent
\begin{equation}
{\vec f}_d=\big( V_X\,\phi_X^{ss},\; V_C\,\phi_C^{ss},\;
-V_X\,\phi_X^{ss\,*},\; -V_C\,\phi_C^{ss\,*}\big)^T
\end{equation}
as well as the induced perturbation $\delta{\vec \phi}_{d} =-{\mathcal
  L}^{-1}\cdot {\vec f}_d$.
The static disorder resonantly excites those Bogoliubov modes
whose frequency is equal to $\omega_p$.
 In the left panels of
  Fig.\ref{fig:comb_coll_k}, the excited modes are given by
  the intersections of the mode dispersion with the horizional
dotted lines. For the specific example of a spatially localized
defect acting on the photonic component, we have plotted in the
right panels of Fig.\ref{fig:comb_coll_k} and in
Fig.\ref{fig:spazio_x} the photonic intensity $|\psi_C|^2$ in
respectively the momentum and the real space for different
parameter regimes. These quantities correspond to the
experimentally accessible far-  and near- field intensity
profiles~\cite{HoudreRRSLin,HoudreRRSNLin,Langbein} of the
resonant Rayleigh scattering of the pump (i.e. the coherently
scattered light at the pump frequency $\omega = \omega_p$). A
similar $\kk$-space pattern is obtained in the presence of a
disordered ensemble of defects.

In the linear regime, the $\kk$-space
emission pattern shown in Fig.\ref{fig:comb_coll_k}(b) contains a peak at the
incident wavevector $\kk_p$, plus the RRS
ring~\cite{HoudreRRSLin,Langbein}. In the real space pattern shown
in  Fig.\ref{fig:spazio_x}(a), as the polariton fluid is moving to
the right, the defect induces a propagating perturbation with
parabolic wavefronts oriented in the left direction.

In the presence of interactions, the RRS circle is transformed
into a $\infty$-like shape with the low-k lobe more intense than
the high-k one. If $\Delta_p<0$, the two lobes are separated by a
gap (Fig.\ref{fig:comb_coll_k}j), while they touch at $\kk_p$ if
$\Delta_p=0$ (Fig.\ref{fig:comb_coll_k}d). In this resonant case,
when $\delta\omega_{MF}$ is large enough for the sound velocity
$c_s$ in the polaritonic fluid to be larger than the flow velocity
$v_p$, the slope of the $+$ branch on the low-k side of the corner
(Fig.\ref{fig:comb_coll_k}e)  becomes positive and there is no
intersection with the horizontal dotted line any longer. In this
regime, RRS is no longer possible, and the polaritonic fluid
behaves as a {\em superfluid} in the sense of the Landau
criterion~\cite{Superfluidity}. Once normalized to the incident
one, the RRS intensity is strongly quenched with respect to the
previous cases and no RRS ring is any longer present. The weak
emission still visible in Fig.\ref{fig:comb_coll_k}(f) is due to
non-resonant processes, which are allowed by the finite broadening
of the polariton modes. As no propagating mode is resonantly
excited, the perturbation in real space remains localized around
the defect, as shown in Fig.\ref{fig:spazio_x}(b). On the other
hand, on the bottom of the bistability curve (where $\Delta_p >
0$), the polariton gas is not superfluid. The RRS intensity is
even enhanced with respect to the linear regime because of the
reduced linewidth of the Bogoliubov modes in the regions where the
$\pm$ branches stick together, as shown in
Fig.\ref{fig:comb_coll_k}(g-h).

To summarize, the polariton fluid has a superfluid behaviour with
respect to static impurities if the equation
$\omega_{LP}^{\pm}(\kk) = \omega_p$ has {\it no} solutions with
$\kk\neq\kk_p$. If the corresponding linear regime equation
$\omega_{LP}(\kk)=\omega_p$ has a set of solutions corresponding
to the elastic RRS ring, the effect of superfluidity is dramatic
as the RRS ring is suppressed. Within the parabolic approximation
in Eq. \eq{Bogo}, a simple {\it sufficient} condition for
superfluidity is found: $\Delta_p \leq 0$ {\em and} $(m_{LP}v_p^2-
\hbar \delta \omega_{MF}) < \hbar  |\Delta_p|$.

In conclusion, we have shown the strict connection between the
dispersion of the elementary excitations in a quantum fluid of
microcavity polaritons and the intensity and shape of the resonant
Rayleigh scattering on defects. In particular, we have pointed out
some experimentally accessible consequences of polaritonic
superfluidity for realistic microcavity parameters. More in
general, thanks to the coupling to externally propagating light,
microcavity polaritonic systems appear to be promising candidates
for the study of novel effects in low-dimensional quantum fluids.

%\section{Acknowledgments}
We acknowledge G. C. La Rocca with whom the original idea of the
work was conceived. We are grateful to Y. Castin and J. Dalibard for
continuous discussions. LKB-ENS and LPA-ENS are two
 "Unit\'es de Recherche de l'Ecole Normale Sup\'erieure et de
l'Universit\'e Pierre et Marie Curie, associ\'ees au CNRS".

\end{document}